\documentstyle[12pt]{article}
\begin{document}
\thispagestyle{empty}

\newcommand{\be}{\begin{equation}}
\newcommand{\ee}{\end{equation}}
\newcommand{\sect}[1]{\setcounter{equation}{0}\section{#1}}
\renewcommand{\theequation}{\thesection.\arabic{equation}}
\newcommand{\vs}[1]{\rule[- #1 mm]{0mm}{#1 mm}}
\newcommand{\hs}[1]{\hspace{#1mm}}
\newcommand{\mb}[1]{\hs{5}\mbox{#1}\hs{5}}
\newcommand{\bea}{\begin{eqnarray}}
\newcommand{\ena}{\end{eqnarray}}
\newcommand{\wt}[1]{\widetilde{#1}}
\newcommand{\und}[1]{\underline{#1}}
\newcommand{\ov}[1]{\overline{#1}}
\newcommand{\sm}[2]{\frac{\mbox{\footnotesize #1}\vs{-2}}
		   {\vs{-2}\mbox{\footnotesize #2}}}
\newcommand{\prt}{\partial}
\newcommand{\eps}{\epsilon}

\newcommand{\R}{\mbox{\rule{0.2mm}{2.8mm}\hspace{-1.5mm} R}}
\newcommand{\Z}{Z\hspace{-2mm}Z}

\newcommand{\cd}{{\cal D}}
\newcommand{\cg}{{\cal G}}
\newcommand{\ck}{{\cal K}}
\newcommand{\cw}{{\cal W}}

\newcommand{\vj}{\vec{J}}
\newcommand{\vl}{\vec{\lambda}}
\newcommand{\vz}{\vec{\sigma}}
\newcommand{\vt}{\vec{\tau}}
\newcommand{\vw}{\vec{W}}
\newcommand{\poiss}{\stackrel{\otimes}{,}}


\newcommand{\NP}[1]{Nucl.\ Phys.\ {\bf #1}}
\newcommand{\PL}[1]{Phys.\ Lett.\ {\bf #1}}
\newcommand{\NC}[1]{Nuovo Cimento {\bf #1}}
\newcommand{\CMP}[1]{Comm.\ Math.\ Phys.\ {\bf #1}}
\newcommand{\PR}[1]{Phys.\ Rev.\ {\bf #1}}
\newcommand{\PRL}[1]{Phys.\ Rev.\ Lett.\ {\bf #1}}
\newcommand{\MPL}[1]{Mod.\ Phys.\ Lett.\ {\bf #1}}
\newcommand{\BLMS}[1]{Bull.\ London Math.\ Soc.\ {\bf #1}}
\newcommand{\IJMP}[1]{Int.\ Jour.\ of\ Mod.\ Phys.\ {\bf #1}}
\newcommand{\JMP}[1]{Jour.\ of\ Math.\ Phys.\ {\bf #1}}
\newcommand{\LMP}[1]{Lett.\ in\ Math.\ Phys.\ {\bf #1}}

\renewcommand{\thefootnote}{\fnsymbol{footnote}}
\newpage
\setcounter{page}{0}
\pagestyle{empty}
\vs{15}
\begin{center}
{\LARGE {\bf On the Super-NLS Equation and its}}\\[.25cm]
{\LARGE {\bf Relation with N=2 Super-KdV }}\\[0.25cm]
{\LARGE {\bf within Coset Approach.}}\\[1cm]

\vs{8}

{\large S. Krivonos$^{a,1}$, A. Sorin$^{a,2}$ and F. Toppan$^{b,3}$}
{}~\\
\quad \\
{\em {~$~^{(a)}$ JINR-Bogoliubov Laboratory of Theoretical Physics,}}\\
{\em 141980 Dubna, Moscow Region, Russia}~\quad\\
{\em ~$~^{(b)}$ Dipartimento di Fisica, Universit\`{a} di Padova,}\\
{\em Via Marzolo 8, 35131 Padova, Italy}

\end{center}
\vs{8}

\centerline{ {\bf Abstract}}

\indent

A manifestly $N=2$ supersymmetric coset formalism is introduced
to describe integrable hierarchies. It is applied to analyze the
super-NLS equation. It possesses an $N=2$ symmetry
since it can be obtained from a manifest $N=2$ coset algebra
construction. A remarkable
result is here discussed: the existence of a
B\"{a}cklund transformation which connects
the super-NLS
equation to the equations belonging to the
integrable hierarchy of one particular (the $a=4$)
$N=2$ super-KdV equation.
$N=2$ scalar Lax pair operators are introduced
for both super-KdV and super-NLS.
\vs{10}
\begin{center}
{\it Submitted to Physics Letters A}
\end{center}
\vfill
\rightline{April 1995}
\rightline{JINR E2-95-185}
\rightline{DFPD 95-TH-24}
\rightline{ hep-th/9504138}
{\em E-Mail:\\
1) krivonos@cv.jinr.dubna.su\\
2) sorin@cv.jinr.dubna.su\\
3) toppan@mvxpd5.pd.infn.it}
\newpage
\pagestyle{plain}
\renewcommand{\thefootnote}{\arabic{footnote}}
\setcounter{footnote}{0}

\sect{Introduction.}

\indent

Studying the hierarchies of integrable differential equations is
an interesting subject not only on a purely mathematical basis but
also, on physical grounds, due to their connection with
the discretized versions
of the two-dimensional gravity.
Looking for supersymmetric extensions of such hierarchies
is therefore quite important because it is commonly believed that
any consistent $2$-dimensional theory which includes gravity
is necessarily
supersymmetric to avoid problems arising from tachyonic states in
the purely bosonic sector. Even if no discretized supergravity
formulation is available at present there has been some attempts
to bypass this step and to formulate such theories directly in terms
of supersymmetric integrable hierarchies.\par
The understanding of bosonic hierarchies is rather good at present
(see \cite{feh}), while the supersymmetric case is
much less understood.\par
In this letter we use an $N=2$ manifestly supersymmetric
formalism to treat with $N=2$ hierarchies and we apply it
to analyze the simplest $N=2$ model, namely the
super-NonLinear Schr\"{o}dinger Equation [2-5].\par
It has been shown in \cite{krisor} that what is commonly
denoted as $N=1$
super-NLS equation [2-4] possesses a hidden $N=2$ supersymmetry
structure. In this letter we will show that such hidden
$N=2$ supersymmetry structure finds a natural explanation
due to the fact that the hierarchy can be obtained as an $N=2$ coset
of an $N=2$ superalgebra.
\par
A remarkable
connection \cite{krisor} between the super-NLS equation and the special
$N=2$ supersymmetric version of KdV obtained by setting
the parameter $a=4 $ (see \cite{mat}) is discussed in the framework of
coset approach.
There exists indeed a B\"{a}cklund transformation which
maps the super-NLS equation into the second flow of the
hierarchy associated to the $N=2$, $a=4$,
KdV equation,
the
super-KdV itself being associated to the third flow.\par
The crucial point which allows recognizing the existence of such
B\"{a}cklund transformation is the existence of
a disguised version \cite{bon} of the bosonic NLS equation which admits
a Virasoro $\times {\hat U(1)}$ Kac-Moody algebra structure;
such equation admits an immediate and natural $N=2$ generalization
by accomodating these fields into a single $N=2$
super stress-energy tensor $J$ (see \cite{krisor}).
\par
The B\"{a}cklund transformation is non-polynomial in the original
fields (and their derivatives) which give rise to the super-NLS equation.
This explains why at a bosonic level (this statement is true also in
the supersymmetric case) a rational ${\cal W}$ algebra structure
associated to the coset model \cite{top1} can be mapped into
a polynomial ${\cal W}$ algebra of the transformed equation.\par
We are able to construct for both the super-NLS equation
and the $N=2$, $a=4$ super-KdV equation
their corresponding
manifestly $N=2$ supersymmetric scalar Lax operators which
generate the infinite tower of hamiltonians in involution.
\par
Particularly interesting is the Lax operator for the super-NLS equation:
even if non-polynomial in the superfields and their derivatives, neverthless
it generates the infinite tower of local (polynomial) hamiltonians
in involution. This fact can be understood
due to the existence of a similarity transformation
which maps the Lax operator into a purely polynomial one.

\section{The $N=2$ formalism and the $N=2$ ${\hat U(2)}$ algebra.}

In this section we will introduce our conventions concerning the
$N=2$ superfield formalism and we will define the $N=2$ algebras
which allow introducing the super-NLS hierarchy.\par
Our superfields $\Phi (x, \theta, {\overline \theta})$ depend on two
conjugate Grassman variables $\theta, {\overline\theta}$.\par
The chiral
and antichiral fermionic derivatives $D, {\overline D}$ are introduced
respectively through
\begin{eqnarray}
D &=& \frac{\partial}{\partial \theta}
-\frac{1}{2}{\overline \theta} \partial_x \; ,\nonumber\\
{\overline D} &=& \frac{\partial}{\partial {\overline \theta}} -
     \frac{1}{2} \theta \partial_x \; .
\end{eqnarray}
They satisfy the following relations
\begin{eqnarray}
D^2 &=& {\overline D}^2 = 0 \; , \nonumber\\
\{ D, {\overline D}\} &=& -\partial_x \; .
\end{eqnarray}
An $N=2$ supersymmetric (bosonic) delta function $\delta (X, Y)$ is
introduced  by assuming
\be
\int dX_1 \delta (X_1,X_2) A(X_1) = A(X_2)
\ee
for any $N=2$ superfield $A(X)$. \\
Here
$X\equiv x, \theta ,{\overline \theta}$
denotes the $N=2$ superspace; the
integration measure is
$dX \equiv dx d\theta d{\overline \theta}$;
the bosonic integration is understood over the
whole real line (or over the circle $S^1$), while the standard
convention for the Berezin integration over
fermionic variables is assumed.\par
Explicitly we have
\be
\delta (X_1, X_2) = \delta (x_1-x_2) (\theta_1 -\theta_2)
({\overline \theta}_1 -{\overline\theta}_2) \; .
\ee

The $N=2$ extension of the ${\hat U(1)}$ Kac-Moody algebra is
introduced through the spin ${\textstyle{1\over 2}}$
fermionic superfields $H, {\overline H}$, satisfying
the following Poisson brackets relations
\begin{eqnarray}
\{ H (1), H (2) \} &=& \{ {\overline H} (1) , {\overline H }(2)\} =0 \; ,
\nonumber\\
\{ H (1) , {\overline H}(2)\} &=& - D_1 {\overline D}_2 \delta ( 1, 2)\; ,
\label{u2al1}
\end{eqnarray}
where, for simplicity, we have just denoted with numbers the superspace
coordinates.\par
Notice that it is consistent with the above relations
to assume that $H, {\overline H}$ are chirally
(and respectively antichirally) constrained superfields, that is
\be
D H = {\overline D}~ {\overline H} = 0
\ee

The notion of chirally covariant fermionic derivatives can now be
introduced in full generality. Let us just remember here that
covariant derivatives have been introduced in \cite{top0} as an
elegant
way to simplify the analysis of $W$-algebras containing Kac-Moody
subalgebras; further it has been shown in \cite{{top1},{top2}} that they
could be useful for constructing integrable hierarchies.
\par
At first we have to precise the notion of charged $N=2$ superfields:
they will be denoted as $\Phi_{q,{\overline q}}$ ($q,{\overline q}$
are respectively the chiral and antichiral charges) and assumed to
satisfy the following Poisson brackets relations with
$H,{\overline H}$:
\begin{eqnarray}
\{H (1) , \Phi_{q,{\overline q}}(2)\} &=& q D_1\delta (1,2)
\Phi_{q,{\overline q}}(2) \; , \nonumber\\
\{ {\overline H} (1) , \Phi_{q,{\overline q}}(2) \} &=&
{\overline q} {\overline D}_1\delta (1,2)
\Phi_{q,{\overline q}}(2) \; .
\label{charge}
\end{eqnarray}
The fermionic covariant derivatives ${\cal D}, {\overline{\cal D}}$
can be introduced through the following definitions:
\begin{eqnarray}
{\cal D} \Phi_{q,{\overline q}} &=& D\Phi_{q,{\overline q}}
-{\overline q} H \Phi_{q,{\overline q}}\; , \nonumber\\
{\overline {\cal D}}\Phi_{q, {\overline q}} &=&
{\overline D} \Phi_{q,{\overline q}}
+q {\overline H} \Phi_{q,{\overline q}} \; , \\
{\cal D}^2 & = &  {\overline {\cal D}}^2 = 0 \; . \nonumber
\end{eqnarray}
They map $\Phi_{q,{\overline q}}$ into new superfields
${\cal D}\Phi_{q,{\overline q}}, {\overline {\cal D}}
\Phi_{q,{\overline q}}$ which still have $q, {\overline q}$
charges, i.e. they satisfy (\ref{charge}).
\par
At this point we can introduce what can be called the $N=2$
extension of the affine ${\hat U(2)}$ algebra \cite{{hull},ASE},
which is obtained
by adding to $H, {\overline H}$ the spin ${\textstyle{1\over 2}}$
fermionic superfields $F, {\overline F}$. They are assumed to be
charged, with charges $(q=1, {\overline q} =-1)$, $(q=-1,
{\overline q} =1)$ respectively.\par
Explicitly we have
\begin{eqnarray}
\{ { H} (1) , F(2) \} &=& { D}_1\delta (1,2) F(2)\; , \nonumber\\
\{ {H} (1) , {\overline  F}(2) \} &=&
- {D}_1\delta (1,2) {\overline F}(2)\; , \nonumber\\
\{ {\overline H} (1) , F(2) \} &=&
-{\overline D}_1\delta (1,2)F(2) \; ,\nonumber \\
\{ {\overline H} (1) , {{\overline F}}(2) \} &=&
 {\overline D}_1\delta (1,2) {{\overline F}}(2) \; . \label{u2al2}
\end{eqnarray}
The algebra is completed with the following relations among
$F,{\overline F}$:
\begin{eqnarray}
\{F(1), F(2) \} &=& \{{\overline F} (1), {\overline F}(2) \} =0 \; ,
\nonumber \\
\{F(1), {\overline F}(2) \}&=&
-2{\cal D }_1 {\overline {\cal D}}_2 \delta (1,2)
+ \delta (1,2) F(2) {\overline {F}}(2) \; ,
\label{u2al3}
\end{eqnarray}
where in the above formulas the compact notation which makes use of
the covariant
derivatives has been introduced, being understood that $\delta(1,2)$
has the required charge properties.\par
The Jacobi identities are satisfied for the above algebra only when
covariantly chiral--anti-chiral constraints for $F, {\overline F} $
are taken into account:
\begin{eqnarray}
{\cal D} F & \equiv & \left( D  + H\right)F= 0 \; ,\nonumber\\
{\overline {\cal D}}\;\!{\overline F}&\equiv& \left( {\overline D}-
             {\overline H} \right){\overline F} =0 \;  .
\end{eqnarray}
The above algebra has the structure of a non-linear
${\cal W}$ algebra when expressed in terms of $N=2$ superfields
due to the presence of non-linear terms in the right hand side.
However, in terms of component fields, the algebra turns out
to be a standard linear algebra. The non-linearity on the right hand side
is the price we have to pay for disposing of a manifestly $N=2$
supersymmetric formalism\cite{{hull},ASE}.\par
It is interesting to notice that in \cite{top2}
the super-NLS equation was derived by coseting the $N=1$
${\hat sl(2)}$ Kac-Moody algebra. Such algebra is not $N=2$
supersymmetric, while its coset has a hidden $N=2$ structure \cite{krisor}.
The algebra here described coincides with an $N=1$ ${\hat sl(2)}\times
{\hat U(1)}$
algebra which possesses a manifest $N=2$ supersymmetry.
The coset itself, which is taken by quotienting the
subalgebra generated by $H,{\overline H}$ is manifestly $N=2$ supersymmetric.
It is associated to a whole super-NLS hierarchy
of integrable equations.
Stated more precisely, there exists an infinite tower
of super-hamiltonian densities which belong to the enveloping
algebra of the $N=2$ ${\hat U(2)}$ algebra and have vanishing
Poisson brackets with respect to $H, {\overline H}$.
The corresponding hamiltonians are all in involution with respect to the
Poisson brackets structure provided by the $N=2$ ${\hat U(2)}$
algebra. \par
Due to their coset structure the whole set of hamiltonian densities
are obtained as
chargeless homogeneous differential polynomials in
$F,{\overline F}$ and the covariant derivatives ${\cal D}, {\overline
{\cal D}}$. The first two hamiltonians are completely specified by
their dimensionality and symmetry properties, while for
higher order hamiltonians, considerations based
on the symmetry properties allows to drastically
reduce their possible form, up to free parameters which must be
fixed by requiring compatibility conditions for the different flows.
Just to give an idea of the simplifications introduced with this formalism,
the third hamiltonian of the hierarchy which will be produced below
is given by the sum of two terms and it is sufficient (up to an overall
normalization factor) to specify one single free parameter.
When the covariant derivatives are expanded
in terms of their $H,{\overline H}$
component superfields, the same hamiltonian turns out to be expressed as
the sum of $23$ different terms.

\section{ The super-NLS hierarchy as $N=2$ coset.}

As discussed in the previous section, there exists an infinite
tower of hamiltonians, all in involution, which belong to the $N=2$
${\hat U(2)}$ coset algebra. This statement will be proved later, with the
introduction of the generating Lax operator.\par
The hamiltonian densities $H_k$, for integers $k=1,2,...$
are bosonic and $k$ is the spin-dimension.
Explicitly we have for the first four hamiltonian densities the
following solutions up to total derivatives:
\begin{eqnarray}
H_1 &=& F{\overline F} \; , \nonumber\\
H_2 &=& F'{\overline F} \; , \nonumber\\
H_3 &=& F'' {\overline F} -\frac{1}{2} {\overline{\cal D}}
 F\cdot{\cal D} {\overline F}\cdot F {\overline F}\nonumber \; , \\
H_4 &=& F'''{\overline F} +\frac{3}{2} {\overline{\cal D}} F\cdot
{\cal D} {\overline F} \cdot F {\overline F}'
+ F' {\overline F}' F{\overline F}  \; , \label{hami}
\end{eqnarray}
where in the above relations we have denoted with a prime
the operator $ -\{ {\cal D}, {\overline{\cal D}}\}$
(i.e. the "covariant" space derivative).\par
The different flows are defined through the position
\begin{eqnarray}
{\textstyle{\partial \over \partial t_k }} \Phi (X)
&=& \{\Phi (X), \int dY H_k (Y) \}
\end{eqnarray}
for any given $\Phi$ superfield. The Poisson brackets are given by
the $N=2$ ${\hat U(2)}$ algebra relations
(\ref{u2al1},\ref{u2al2}-2.11).\par
The first flow just gives the covariantly chiral equations of motion
for $F,{\overline F}$
\begin{eqnarray}
{\textstyle{\partial\over \partial t_1}} F &=& F~' \; , \nonumber\\
{\textstyle{\partial\over\partial t_1}} {\overline F} &=&
{\overline F}~' \; ,
\end{eqnarray}
while the second hamiltonian is the one which
gives the $N=2$ super-NLS \cite{krisor}:
\begin{eqnarray}
{\textstyle{\partial\over\partial t_2}} F&=&
F~'' -F {\cal D}({\overline F}\;\! {\overline{\cal D}} F),
\nonumber\\
{\textstyle{\partial\over \partial t_2}} {\overline F} &=&
-{\overline F}~'' +{\overline F}\;\! {\overline{\cal D}}
(F{\cal D}{\overline F}) \; .
\label{supnls}
\end{eqnarray}
As for the third flow we have explicitly
\begin{eqnarray}
{\textstyle{\partial\over\partial t_3}}
F &=& F~''' -\frac{3}{2}\left[ ({\overline{\cal D}}F~')F{\cal D}{\overline F}
+F~' ({\overline {\cal D}} F) {\cal D} {\overline F}
+ F~'' F {\overline F} \right] \; , \nonumber\\
{\textstyle {\partial\over\partial t_3}} {\overline F} &=& {\overline F}~'''-
\frac{3}{2}\left[ ({\cal D} {\overline F}~' ) {\overline F}\;\!
{\overline {\cal D } } F
+{\overline F}~' ({\cal D}{\overline F}){\overline{\cal D}} F
+{\overline F} ~''{\overline F} F\right] \; .
\label{three}
\end{eqnarray}
As we will show in the next section these equations are related with
$N=2$ super-KdV equation \cite{mat} via B\"{a}cklund transformation.

Since all the hamiltonian densities $H_k$ have vanishing Poisson brackets
with the superfields $H, {\overline H}$, then
\begin{eqnarray}
{\textstyle{\partial\over\partial t_k}} H &=&
{\textstyle{\partial\over\partial t_k}} {\overline H} =0
\end{eqnarray}
for any $k$.\par
It is therefore consistent to set, at the level of the equations of motion,
$H={\overline H} =0$, which implies in particular
that the covariant
derivatives in the expressions for $F, {\overline F}$ can be replaced
by the ordinary fermionic derivatives. \par
The formalism of coset here used allows a simpler analysis than the
procedure based on
Dirac's constraints: $H={\overline H}=0$ is obtained as a consequence
of the equations of motion and not imposed as an external constraint;
it is therefore not necessary to use Dirac's brackets. This simplification
is very much evident when using $N=2$ superfield formalism since the
algebra between $H, {\overline H}$ (\ref{u2al1}) involves
a non-invertible operator ($D{\overline D}$) acting on the delta-function,
which puts extra-complications in the Dirac procedure
in this case.

\section{The B\"{a}cklund transformation bewteen super-NLS and super-KdV.}

At the bosonic level it is very well known that there exists a
mapping between the
fields $J_\pm$ entering the standard NLS equation and the fields
$R,S$ of ref. \cite{bon} which satisfy an equation equivalent to the NLS one.
The $N=2$ generalization of such mapping \cite{krisor} is provided
by the following B\"{a}cklund transformation,
local but not polynomial in the
fields $F, {\overline F}$ and their covariant derivatives, mapping
them into a single spin $1$ $N=2$ superfield $J$:
\begin{eqnarray}
J&=&  \frac{1}{4} {\overline F} F - \frac{1}{2}{\textstyle {D}
{\overline F}~'\over {D}{\overline F}} \; .
\label{back}
\end{eqnarray}
Notice that this transformation is not invertible.
\par
On $J$ the reality condition $ J = J^*$ can be imposed.
\par
A feature which is absolutely new in the super-case and is not
present in the bosonic case is the following: the above
B\"{a}cklund transformation connects two very well studied hierarchies,
the super-NLS one with the $a=4$ $N=2$ super-KdV \cite{mat}.
In the bosonic case
the KdV equation is obtained only after performing a field reduction
\cite{bon}, while here we have an exact equivalence (we recall that
$N=2$ super-KdV contains one extra-field in the bosonic sector with
respect to the standard KdV).\par
In \cite{{bdas2},{bdas3}} it has been shown that the $N=1$ super-KdV can be
obtained after a suitable reduction of the $N=1$ super-NLS
equation, but the
exact equivalence
with the $N=2$ super-KdV has not been remarked.\par
It can be easily realized that the hamiltonians (\ref{hami})
of the super-NLS hierarchy, and the corresponding flows
for $F, {\overline F}$,
\footnote{Here we assume for simplicity the reduction $H={\overline H} =0$,
but the general case works as well by substituting the
fermionic derivatives
with the covariant ones} can be reobtained respectively
from the hamiltonians
of the $N=2$ $a=4$ super-KdV hierarchy and the corresponding flows for $J$,
after reexpressing $J$ in terms of (\ref{back}).
\par
We have, as first order hamiltonian densities for such a hierarchy \cite{mat}:
\begin{eqnarray}
{\cal H}_1 &=& J \; , \nonumber\\
{\cal H}_2 &=& J^2 \; , \nonumber\\
{\cal H}_3 &=& J^3 +\frac{3}{4} J[D, {\overline D}] J
 \; , \nonumber \\
{\cal H}_4 &=& J^4 -\frac{1}{2} (J~')^2 +\frac{3}{2}
J^2 [D,{\overline D} ] J \; .
\label{hamij}
\end{eqnarray}
The second Poisson brackets structure is just given by the $N=2$
superconformal algebra; explicitly we have in the $N=2$ superfield
formalism
\begin{eqnarray}
\{ J(X_1), J(X_2) \} &=&
({\textstyle{1\over 2}}
[D, {\overline D}]\partial + J(X_2) \partial + ({\overline D}J(X_2))D
   \nonumber\\
  &&+ (D J(X_2)) {\overline D}
+ J(X_2)~') \delta(X_1,X_2)
\end{eqnarray}
where all derivatives in the above formula are applied in $X_2$.\par
The following flows can be derived, with a suitable overall
normalization for the hamiltonians in (\ref{hamij}):
\begin{eqnarray}
{\textstyle{\partial\over\partial t_1}} J &=& J~' \; , \nonumber\\
{\textstyle{\partial\over\partial t_2}} J &=& [D, {\overline D}]
J~' + 4 J~' J \; , \nonumber\\
{\textstyle{\partial\over\partial t_3}} J &=&
- J~''' + 6 ({\overline D} J D J)~'
-6  (J [D, {\overline D}] J)~' -4 (J^3)~' \; .
\end{eqnarray}
The second flow corresponds precisely to the equation
derived from the super-NLS equation
once the B\"{a}cklund transformation (\ref{back})
has been taken into account,
while the third flow is just the $a=4$
(in the language of \cite{mat}) $N=2$ super-KdV equation.\par
The third flow (\ref{three}) generates for the B\"{a}cklund
transformed superfield $J$ just the above super-KdV equation.
\par
It should be noticed that the B\"{a}cklund transformation (\ref{back})
connects through field redefinitions the two hierarchies of
equations of motion, but it cannot be lifted to a Poisson map relating the
two Poisson structures, the one given by the $N=2$ ${\hat {U(2)}}$
algebra, with the one given by $N=2$ super-Virasoro.
This means in particular that (\ref{back}) is
{\em not}
a Sugawara-type realization of the $N=2$ super-Virasoro
superfield $J$; this statement
holds even if we replace the standard fermionic derivative with the covariant
one in (\ref{back})).

\section{The Lax pairs.}

In this section we construct new
bosonic manifestly $N=2$ supersymmetric Lax pairs for
the super-NLS hierarchy and the super-KdV hierarchies.
\par
At first we will compute the Lax operator $L$ in terms of the $J$
superfield. It can be constructed by assuming as an Ansatz
that it has a non-standard form (see \cite{kup}), already introduced
in a slightly different context and for a different Lax operator in
\cite{popow}: the different flows are defined through
the equation
\begin{eqnarray}
{\textstyle{\partial\over\partial t_k}}L &=&
[ (L^k)_{\geq 1} , L]
\label{lax1}
\end{eqnarray}
for integer values of $k=1,2,...$ .
The underscript $\geq 1$ means that
only the purely derivative part must be considered \cite{popow}.  \par
In our case the integrals of motion $I_n$ are obtained from the constant
term of $L^n$, that is
\begin{eqnarray}
{H}_n &=& I_n = ({-{\textstyle {1\over 2}}})^n\int d X (L^n)_{0} \; ,
\end{eqnarray}
where underscript $0$ means the constant part of an operator.
They coincide with the integrals of the
hamiltonian densities given in (\ref{hamij}).
This statement has been explicitly checked, with computer computations,
up to the fourth order hamiltonian of the series.\par
It should be noticed that our prescription to compute the
integrals over the residue is not the canonical one. In the standard
prescription for the
$N=2$ formalism the integrals are computed from the term
$[D, {\overline D}]\partial^{-1}$. For our Lax operator
all these integrals are vanishing. The only
possibility left
to compute integrals of motion
which is consistent due to dimensional reasons
and the bosonic character of the $N=2$ integration measure
is precisely the one here given. Our computations prove
that this is indeed so. We believe that our Lax operator is
the first representative of a new class of
manifestly $N=2$ supersymmetric Lax operators
admitting such property.
\par
The explicit expression for
our Lax operator which provides
the flows for the $N=2$ $a=4$ super-KdV hierarchy is given by the
following formula:
\begin{eqnarray}
L&=& \partial -2J -2{\overline D} \partial^{-1} (DJ)\; .
\end{eqnarray}
Here $\partial$ means the standard bosonic $x$-derivative. The brackets
in the above expression means the action of $D$ on $J$, (that is
$D$ must not be commuted with $J$).\par
Inserting the B\"{a}cklund transformation (4.18) in the above Lax pair
allows us to produce, with the same prescription as before, the
integrals of motion for the super-NLS hierarchy. We have
in this case
\begin{eqnarray}
L&=& \partial - \frac{1}{2}{\overline F} F +
   \frac{D {\overline F}'}{D {\overline F}}
- \frac{1}{2}{\overline{D}}\partial^{-1} F ({D {\overline F}})\; .
\end{eqnarray}
One should immediately notice that the above Lax pair is not polynomial
in the superfields $F, {\overline F}$ and their derivatives.
Despite of this, it is quite remarkable that the conserved
quantities which are produced are local, polynomial expressions
which coincides, of course, with the hamiltonians already computed
for the super-NLS hierarchy. \par
The arising of polynomial hamiltonians can be
clearly understood once realized that there exists
a similarity transformation which maps $L$ into an operator
${\tilde L}$ through
\begin{eqnarray}
L& =& \frac{1}{( D{\overline F})} {\tilde L} (D{\overline F}) \; , \\
{\tilde L} & = & \partial -\frac{1}{2}{\overline F}F -
   \frac{1}{2}(D{\overline F}){\overline D}\partial^{-1}F \; . \nonumber
\end{eqnarray}
${\tilde L}$ does not produce consistent equations of motion, but
its formal adjoint operator ${\tilde L}^*$, given by
\begin{eqnarray}
{\tilde L}^* &=& \partial +\frac{1}{2}{\overline F} F -
 \frac{1}{2} F {\overline D} \partial^{-1}(D {\overline F})
\end{eqnarray}
turns out to be a polynomial Lax operator which provides
the flows and conserved integrals of motion for the super-NLS hierarchy
(with the same conventions as before).\par
In \cite{bdas3} analogous steps (that is at first taking
a similarity transformation
and then the formal adjoint of the transformed operator)
were introduced to produce a polynomial Lax pair operator
for the super-NLS hierarchy out of a starting non-polynomial one.
The Lax operators here produced are different from those of
\cite{bdas3} which are expressed in terms of $N=1$ superfields.

\section{Conclusions.}

In this paper we considered a remarkable and unexpected
connection \cite{krisor} between
the super-NLS equation and the $N=2$ super-KdV hierarchy
in the framework of coset approach.
The deep reason for the existence of the B\"{a}cklund transformation
relating the two hierarchies is still mysterious to us. However it sheds
light on the fact that hierarchies which are produced in completely
different manners and apparently have nothing in common can be intimately
related. This provides further motivation for looking at other such kinds
of relations.\par
Some problems are more naturally studied in one framework rather than
the other one. So, for instance, Lax pairs are more easily
obtainable in the context of the super-KdV,
while the coset derivation is very transparent in the super-NLS
language.\par
We plan in future to further develop our
$N=2 $ formalism to study and produce more complicated $N=2$
hierarchies.

\section*{Acknowledgements.}

We are indebted to L. Bonora, E. Ivanov and A. Pashnev for interesting
discussions. One of us (F.T.) wishes to express its gratitude to
the Director of Bogolyubov Theor. Lab. of JINR for the kind
hospitality in Dubna where this work has been written.

This investigation has been supported in part by the Russian Foundation of
Fundamental Research, grant 93-02-03821, and the International Science
Foundation, grant M9T300.

\end{document}